\documentclass[12pt]{article}
\usepackage{color}
\usepackage{latexsym}
\usepackage{epsfig,amssymb,euscript, mathrsfs}
\usepackage{amsmath}
\usepackage{graphicx}
\newsavebox{\ns}
\newsavebox{\dbrane}
\newsavebox{\dbshort}

\def\be{\begin{equation}}
\def\ee{\end{equation}}
\def\bea{\begin{eqnarray}}
\def\eea{\end{eqnarray}}

\newcommand\Z{\mathbb{Z}}

\newcommand\C{\mathbb{C}}

\newcommand\diff{\mathrm{d}}

\newcommand{\dd}{\mathrm{d}}

\newcommand{\ii}{\mathrm{i}}

\newcommand{\ex}{\mathrm{e}}

\newlength{\sswidth}

\newcommand{\bm}{\mathbf{m}}
\newcommand{\al}{z}
\newcommand{\tbm}{\mathbf{\tilde{m}}}
\newcommand{\as}{\mathrm{asymp}}
\newcommand{\xmin}{x_{\mathrm{min}}}
\newcommand{\xmax}{x_{\mathrm{max}}}

\numberwithin{equation}{section}       

\begin{document}

\begin{titlepage}

\begin{center}

\today

\vskip 2.3 cm 

{\Large \bf The large $N$ limit of M2-branes}
\vskip .5cm 
{\Large \bf on Lens spaces}

\vskip 2 cm

{Luis F. Alday$^1$, Martin Fluder$^2$ and James Sparks$^1$\\}

\vskip 1cm

$^1$\textit{Mathematical Institute, University of Oxford,\\
24-29 St Giles', Oxford OX1 3LB, United Kingdom\\}

\vskip 0.8cm

$^2$\textit{Rudolf Peierls Centre for Theoretical Physics, \\
University of
  Oxford, 1 Keble Road, Oxford OX1 3NP, U.K.}

\end{center}

\vskip 2 cm

\begin{abstract}
\noindent We study the matrix model for $N$ M2-branes wrapping a Lens space $L(p,1)=S^3/\Z_p$. 
This arises from localization of the partition function of the ABJM theory, and 
has some novel features compared with the case of a three-sphere, including a sum over
flat connections and a potential that depends non-trivially on $p$. We study the matrix model
both numerically and analytically in the large $N$ limit, finding that a certain family of $p$ flat connections 
give an equal dominant contribution. At large $N$ we find the same eigenvalue distribution for all $p$, 
and show that the free energy is simply $1/p$ times the free energy on a three-sphere, 
in agreement with gravity dual expectations. 
 \end{abstract}

\end{titlepage}

\pagestyle{plain}
\setcounter{page}{1}
\newcounter{bean}
\baselineskip18pt
\tableofcontents

\section{Introduction and summary}

One of the most exciting features of supersymmetric gauge theories is that one can compute certain protected quantities 
exactly. The most fundamental of these quantities is the partition function. In recent years 
localization techniques have been developed that allow the computation of partition functions 
for supersymmetric gauge theories in different dimensions, and on different backgrounds, starting with the work of \cite{Pestun:2007rz, Kapustin:2009kz}. 
These results have also led to new tests of conjectured dualities between theories.

Three-dimensional $\mathcal{N}=2$ gauge theories form a particularly fertile ground in which to develop these ideas. 
The simplest compact manifold on which one can define $d=3$ supersymmetric field theories is the round three-sphere, 
originally studied in \cite{Kapustin:2009kz}. This generalizes to other three-manifolds $M_3$, including 
$S^1\times S^2$ and certain one-parameter families of squashed three-spheres \cite{Hama:2011ea, Imamura:2011wg}. 
In this paper 
we study $\mathcal{N}=2$ theories on the Lens spaces $L(p,1)$, which are free
quotients of $S^3$ by $\Z_p$. 
In the first part of the paper we  derive a formula for the full localized partition
 function of a three-dimensional $\mathcal{N}=2$ $U(N)$
Chern-Simons-matter theory on such a Lens space. Several of the ingredients 
have already appeared in previous papers, including
 \cite{Gang:2009wy, Kallen:2011ny, Benini:2011nc}. The 
partition function reduces to a matrix model integral, 
where the potential function depends non-trivially on $p$.

In the second part of the paper we consider the partition function 
in the large $N$ limit, keeping the Chern-Simons levels fixed. 
The motivation for this is that, for appropriate matter content, 
one expects to be able to reproduce these results 
from a dual M-theory gravity computation. 
In particular, we focus on the low energy 
effective theory on $N$ M2-branes, described by the ABJM 
theory \cite{Aharony:2008ug}.
A new feature that arises when $M_3$ has  non-trivial fundamental group 
is that one must sum over different topological sectors in the partition function. 
In the present case, different sectors are labelled by a diagonal $N\times N$ matrix
with entries in $\Z_p$. In the large $N$ limit, we show that one can in fact 
focus on the contribution from matrices proportional to the identity. 
This drastically simplifies the analysis.

At large $N$ we may use a saddle point approximation 
to the matrix model, following \cite{Herzog:2010hf}. 
The leading contribution to the free energy arises from a specific 
eigenvalue distribution. In order to gain some intuition 
we study this distribution numerically, for a number of 
values of $p$. These numerical results lead to a simple ansatz for the eigenvalue 
distribution at large $N$. We then use this ansatz to obtain 
analytic results for the free energy, as well as for the eigenvalue 
distribution and corresponding density. 
We find that the eigenvalue behaviour is in fact independent of 
$p$, with a free energy that is simply $1/p$ times the free energy on a three-sphere. 
This is in agreement with gravity dual expectations, where $L(p,1)$ arises 
as the conformal boundary of AdS$_4/\Z_p$. 

The organization of this paper is as follows. In section \ref{sec:localize} we 
derive the full localized partition function for a  three-dimensional $\mathcal{N}=2$ 
Chern-Simons-matter theory on the Lens space $L(p,1)$. Section \ref{sec:numerics} 
contains the numerical results, which are compared to 
corresponding analytic results in section \ref{sec:analytic}. We 
mention some open problems in the outlook section \ref{sec:outlook}. 
Finally, some technical results are relegated to the
appendices.

\section{The localized partition function on $L(p,1)$}\label{sec:localize}

In this section we derive a formula (\ref{partitionfunction}) for the full localized partition function of a three-dimensional $\mathcal{N}=2$ 
Chern-Simons-matter theory on the Lens space $L(p,1)=S^3/\Z_p$. 
We then specialize to the ABJM theory \cite{Aharony:2008ug} on $N$ M2-branes of interest.

\subsection{The Lens space $L(p,1)$}\label{sec:Lens}

The Lens space $L(p,1)=S^3/\Z_p$ is a certain freely-acting quotient of the round $S^3$ by 
a group of order $p$. Regarding $S^3$ as a unit sphere in Euclidean $\C^2$, with complex 
coordinates $z_1,z_2$, the $\Z_p$ action is generated by
\bea\label{Zpaction}
\C^2 \ \ni \ (z_1,z_2) \ \mapsto \ ( \omega_p\,  z_1, \omega_p^{-1} z_2)~, 
\eea
where $\omega_p=\ex^{2\pi \ii/p}$ is a primitive $p$-th root of unity. Notice this is simply a $\Z_p$ quotient along 
the $S^1$ fibre of the Hopf fibration: $S^1\hookrightarrow S^3\rightarrow S^2$.

Quotienting $S^3$ by the free action (\ref{Zpaction}) leads to a smooth three-manifold $L(p,1)$ 
with $\pi_1(L(p,1))\cong \Z_p$. There are then $p$ topologically inequivalent complex line bundles $L$
over $L(p,1)$, labelled by their first Chern class $c_1(L)\in H^2(L(p,1);\Z)\cong \Z_p$. 
Each $L$ admits a flat $U(1)$ connection, which plays an important role 
in studying gauge theory on $L(p,1)$. For example, rather than complex-valued
functions on $L(p,1)$, it will be important to consider more generally sections of 
$L$. 

Concretely, one can construct such sections as certain projections 
of functions on the covering space $S^3$. For example, we may 
expand complex-valued functions on $S^3$ in terms of hyperspherical harmonics
\bea\label{hyperspherical}
Y_{\ell,m,n}(\theta,\phi,\psi) &=& y_{\ell,m,n}(\theta)\ex^{\ii m\phi} \ex^{\ii n\psi}~,
\eea
where $(\theta,\phi,\psi)$ are standard Euler angles on $S^3$. 
Here $\ell\in\Z_{\geq 0}$ while $m,n\in \{-\frac{\ell}{2},-\frac{\ell}{2}+1,\ldots,\frac{\ell}{2}\}$, 
with $\ell$ labelling the $(\frac{\ell}{2},\frac{\ell}{2})$ spin representation of 
$SU(2)_L\times SU(2)_R$ acting on $S^3$. For example, $Y_{\ell,m,n}$ 
has eigenvalue $-\ell(\ell+2)$ under the Laplacian. 

If $z_i=r_i\ex^{\ii \varphi_i}$ denote 
polar coordinates on each copy of $\C$ in $\C^2$, 
then $\phi=\varphi_1+\varphi_2$, $\psi=\varphi_1-\varphi_2$, and 
in terms of Euler angles 
the generator (\ref{Zpaction}) thus acts as
\bea\label{Euleraction}
\psi & \mapsto & \psi + \frac{4\pi}{p}~.
\eea
The complex-valued functions on $L(p,1)$ are precisely 
the $\Z_p$-invariant functions on $S^3$, and thus (\ref{Euleraction}) 
and (\ref{hyperspherical}) imply that functions on 
$L(p,1)$ are spanned by the modes (\ref{hyperspherical}) satisfying
\bea
2n &\equiv & 0 \ \mbox{mod}\ p~.
\eea
More generally,  sections of $L$ are spanned by the modes (\ref{hyperspherical})
satisfying
\bea\label{projection}
2n &\equiv & c_1(L) \ \mbox{mod}\ p~,
\eea
where we are using the isomorphism $H^2(L(p,1);\Z)\cong \Z_p$. 
This follows since the holonomy of the flat connection $A$ on $L$ around 
the generator $\gamma$ of $\pi_1(L(p,1))$ is
\bea
\exp\left[\ii \int_\gamma A\right] &=& \ex^{2\pi \ii c_1(L)/p}~.
\eea
Here $\gamma$ is represented by a circle fibre in $S^1\hookrightarrow L(p,1)\rightarrow S^2$.
Sections of $L$ must then also pick up this phase around $\gamma$.

Another issue, important for considering supersymmetric field theories, 
concerns the Killing spinors. The 4 Killing spinors on $S^3$ transform in the 
$(\mathbf{2},\mathbf{1})\oplus (\mathbf{1},\mathbf{2})$ representation 
of $SU(2)_L\times SU(2)_R\cong \mathrm{Spin}(4)$. 
The spinor used for localization in \cite{Kapustin:2009kz} 
is in the $(\mathbf{2},\mathbf{1})$ representation. 
In this language the $\Z_p$ action in (\ref{Euleraction}) 
is contained in $U(1)_R\subset SU(2)_L\times SU(2)_R$, 
and hence the Killing spinor used for localization on $S^3$ projects down to a Killing 
spinor on $L(p,1)$. 

We pause here to make some comments on more
 general Lens spaces $L(p,q)$, with $q>1$.
 These are defined 
as the free quotient of $S^3\subset\C^2$ by the action
\bea\label{Zpqaction}
\C^2 \ \ni \ (z_1,z_2) \ \mapsto \ ( \omega_p^q\,  z_1, \omega_p^{-1} z_2)~, 
\eea
with $q$ relatively prime to $p$. The $\Z_p$ action (\ref{Euleraction}) now becomes
\bea\label{Eulerpqaction}
\phi & \mapsto & \phi + \frac{2\pi(q-1)}{p}~,\nonumber\\
\psi & \mapsto & \psi + \frac{2\pi(q+1)}{p}~,
\eea
on the Euler angles.
In particular, there is therefore no invariant spinor, unless $q=1$ or $q=p-1$. 
This makes the treatment for this case more involved, and we therefore leave it for future work.

\subsection{Localization of the path integral}

The localization of the path integral on $L(p,1)$ is very similar to the 
original computation for $S^3$ in \cite{Kapustin:2009kz}. Indeed, 
locally the spaces are identical, so one just needs to keep track of 
how global differences affect formulae. For example, for 
a $U(N)$ gauge theory the path integral still localizes onto flat 
connections $A$, but on $L(p,1)$ there are non-trivial flat connections 
that one must then sum over. A flat $U(N)$ connection on a manifold $M$ is determined by its holonomies,
which define a homomorphism $\varrho:\pi_1(M)\rightarrow U(N)$. Gauge transformations 
act by conjugation, so that flat $U(N)$ connections are in 1-1 correspondence with 
\bea
\mathrm{Hom}(\pi_1(M)\rightarrow U(N))/\mathrm{conjugation}~.
\eea
Since $\pi_1(L(p,1))\cong\Z_p$,  specifying 
a flat connection is equivalent to specifying 
the holonomy around 
the generator $\gamma$ of $\pi_1(L(p,1))\cong \Z_p$
\bea
\mathrm{hol}_\gamma\left(A\right) &=& \mathrm{diag}\left(\omega_p^{m_1},\ldots,\omega_p^{m_N}\right)~,
\eea
where $0\leq m_i <p$, and $i=1,\ldots,N$ runs over the generators of the Cartan 
 $U(1)^N$ subgroup  of $U(N)$. Here we order $m_1\leq m_2\leq \ldots\leq m_N$ 
(conjugation permutes the entries).
The localized path integral will then 
give a sum over topological sectors $\mathbf{m}=\mathrm{diag}(m_1,\ldots,m_N)$, for each 
$U(N)$ gauge group.

Apart from this, as for $S^3$ all fields localize to zero except for the 
D-term and scalar $\sigma$ in the $\mathcal{N}=2$ vector multiplet, which are related via
\bea\label{Dterm}
D &=& -\sigma~.
\eea
The scalar $\sigma$ must be covariant constant. Writing the flat gauge field defined by 
$\bm$ as $A_\bm=-\ii g_\bm^{-1}\diff g_\bm$, this implies that $\sigma=g_\bm^{-1}\sigma_0
g_\bm$ where $\sigma_0$ is a \emph{constant} 
$N\times N$ Hermitian matrix satisfying
\bea
[\sigma_0,\bm]&=& 0~.
\eea
 For a Chern-Simons gauge theory, this saddle point solution gives a standard classical contribution 
to the saddle point approximation of the path integral
\bea\label{classical}
\exp\left[-S_{\mathrm{classical}}(\sigma_0)\right] &=& \exp\left[\frac{\ii \pi k}{p}\mathrm{Tr} (\sigma_0^2)\right]~,
\eea
coming from the supersymmetric completion of the Chern-Simons interaction, evaluated on (\ref{Dterm}). Here 
$k\in \Z$ is the Chern-Simons level. The $p$-dependence in (\ref{classical})
 simply arises because $\mathrm{Vol}(L(p,1))=\mathrm{Vol}(S^3)/p$.
The path integral then reduces to a matrix integral over $\sigma_0$, as well as the discrete sum over $\mathbf{m}$ labelling flat 
$U(N)$ gauge fields.
One must also include the Chern-Simons action for the flat 
gauge field:
\bea\label{flatCS}
\exp\left[-S_{CS}(A)\right] & = &  \exp\left[-\frac{\ii k}{4\pi}\int_{L(p,1)} \mathrm{Tr}\left(A\wedge \diff A+\frac{2}{3}A^3\right)\right]\nonumber\\
&=& \exp\left[-\frac{\ii\pi k }{p} \mathrm{Tr}\, (\mathbf{m}^2)\right]~.
\eea
One computes (\ref{flatCS}) in a standard way: choose a four-manifold $M_4$
with boundary $\partial M_4 = L(p,1)$, and an extension of the bundle and (flat) connection 
$A$ on $L(p,1)$ to corresponding data over $M_4$. 
The Chern-Simons action is then in fact \emph{defined} as $-(\ii k/4\pi)\int_{M_4}\mathrm{Tr}\, (F\wedge F)$, 
which can be shown to be independent of choices, modulo $2\pi \ii$. 
For example, in the present case one can take $M_4=$ total space of $\mathcal{O}(p)\rightarrow\mathbb{CP}^1$, and 
note that the restriction map $\Z\cong H^2(M_4;\Z) \rightarrow H^2(L(p,1);\Z)\cong\Z_p$ is simply 
reduction mod $p$. 

Having summarized the localization, we next turn to the effect on the one-loop 
contributions around the saddle points specified by $(\sigma_0,\mathbf{m})$. 
Due to the remarks in  section \ref{sec:Lens}, the spectra of operators  that 
contribute to the one-loop determinants reduce to an appropriate projection 
of the full spectra on $S^3$.

\subsection{Matter multiplet}

We consider here the contribution of a chiral matter field $\Phi$, in the representation $\mathcal{R}$ of the gauge group, 
to the one-loop determinant around the classical background labelled by $(\sigma_0,\mathbf{m})$. 
We denote the R-charge of $\Phi$ as $\Delta=\Delta(\Phi)$ -- the canonical 
value is $\Delta=\frac{1}{2}$ --  and the weights of $\mathcal{R}$  by $\rho$. 

The bosonic contribution to the one-loop determinant is then \cite{Kapustin:2009kz, Jafferis:2010un, Hama:2010av}
\bea\label{boson}
\mathrm{det}_{\ell/2,m}(D_{\mathrm{boson}}) &=& \ell(\ell+2)- 4m(1-\Delta)+\Delta^2+\rho(\sigma_0)^2~,
\eea
where $\ell,m$ label the same quantum numbers as in section \ref{sec:Lens}, 
so that $\ell\geq 0$ and $m\in\{-\frac{\ell}{2},-\frac{\ell}{2}+1,\ldots,\frac{\ell}{2}\}$. In particular, 
the $\ell(\ell+2)$ term simply comes from the eigenvalue under (minus) the scalar Laplacian. 
On $S^3$ there are $\ell+1$ such modes, labelled by the quantum number $n\in\{-\frac{\ell}{2},-\frac{\ell}{2}+1,\ldots,\frac{\ell}{2}\}$, 
while on $L(p,1)$ we should keep only those modes satisfying 
\bea\label{matterprojection}
2n &\equiv & \rho(\bm) \ \mbox{mod}\ p~,
\eea
as follows from (\ref{projection}).

The fermionic contribution to the one-loop determinant is also given by (\ref{boson}), but now with 
$m\in\{-\frac{\ell}{2},-\frac{\ell}{2}+1,\ldots,\frac{\ell}{2}-1\}$, and with 
an additional contribution of $(-1)^\ell \left(\ell+\Delta+\ii\rho(\sigma_0)\right)\left(\ell+2-\Delta+\ii\rho(\sigma_0)\right)$.
Again, on $S^3$ there is a degeneracy of $\ell+1$, labelled by $n$, while on $L(p,1)$ 
we should keep only those modes satisfying (\ref{matterprojection}).

Since the one-loop determinant is a ratio of fermionic and bosonic determinants, 
we thus see that for fixed $\ell, m$ and $\mathbf{m}$, for every choice of $n$ satisfying (\ref{matterprojection}) 
the contributions from fermionic and bosonic determinants will cancel, \emph{except} 
for the ``missing'' fermionic mode with $m=\tfrac{\ell}{2}$ -- this remains uncancelled 
in the bosonic determinant. We thus conclude that, for fixed $\ell$, we have
\bea
\frac{\mathrm{det}_{\ell/2}(D_{\mathrm{fermion}})}{\mathrm{det}_{\ell/2}(D_{\mathrm{boson}})} &=& 
(-1)^\ell\frac{ \left(\ell+\Delta+\ii\rho(\sigma_0)\right)\left(\ell+2-\Delta+\ii\rho(\sigma_0)\right)}{ \ell(\ell+2)- 2\ell(1-\Delta)+\Delta^2+\rho(\sigma_0)^2}\nonumber\\
&=&  (-1)^\ell \frac{\ell + 2-\Delta+\ii\rho(\sigma_0)}{\ell+\Delta-\ii\rho(\sigma_0)}~,
\eea
where the degeneracy is the number of half-integers $n\in\{-\frac{\ell}{2},-\frac{\ell}{2}+1,\ldots,\frac{\ell}{2}\}$
satisfying
\bea
2n &\equiv & \rho(\mathbf{m}) \ \mbox{mod} \ p~. 
\eea
This degeneracy was denoted by $N_\rho(\ell)$ in \cite{Benini:2011nc}. Thus in total
\bea\label{matter}
Z^{\mathrm{matter}}_{\mathrm{1-loop}}(\sigma_0,\bm) &=& \prod_{\rho\in\mathcal{R}}\prod_{\ell\geq 0}\left(\frac{\ell+2-\Delta+\ii\rho(\sigma_0)}{\ell+\Delta-\ii
\rho(\sigma_0)}\right)^{N_\rho(\ell)}~.
\eea

\subsection{Vector multiplet}

The analysis for the one-loop contribution of the vector multiplet is very similar. 
Here it is more convenient to follow the analysis in \cite{Hama:2011ea}, 
where rather than working out the full spectrum, most of which then 
cancels in the ratio of determinants, instead one isolates the uncancelled 
modes from the outset. These are precisely the eigenmodes which are not paired 
with a superpartner. We shall refer to these as the uncancelled modes.

The uncancelled gaugino modes on $S^3$ have eigenvalues 
\bea\label{mus}
\mu &=& n_1+n_2+\ii\alpha(\sigma_0)
\eea
under the relevant Dirac operator, where $n_i$ denote the charges under 
$\partial_{\varphi_i}$, where recall that $\varphi_1$, $\varphi_2$ are 
azimuthal angles on each copy of $\C$ in $\C^2\supset S^3$.
The normalizable modes are $\{n_1,n_2\geq 0\}\setminus\{(n_1,n_2)=(0,0)\}$. 
The corresponding charges under $\partial_\psi$, $\partial_\phi$ 
are then $\frac{1}{2}(n_1-n_2)$, $\frac{1}{2}(n_1+n_2)$, 
respectively (see just before equation (\ref{Euleraction})), so that the projection condition becomes
\bea\label{niequiv}
n_1 &\equiv & n_2 + \alpha(\bm) \ \mathrm{mod} \ p~.
\eea
The uncancelled transverse vector modes also have eigenvalues of the form 
(\ref{mus}), except now $n_1,n_2\leq -1$. Thus 
\bea
Z^{\mathrm{vector}}_{\mathrm{1-loop}}(\sigma_0,\bm) \ =\ \prod_\alpha\left[ \frac{{\displaystyle\prod}_{\mbox{\tiny $\begin{array}{c}n_1,n_2\geq0 \\ (n_1,n_2)\neq (0,0) \\ n_1\equiv n_2 + \alpha(\bm) \ \mbox{mod}\ p\end{array}$}} [n_1+n_2+\ii\alpha(\sigma_0)]}{{\displaystyle\prod}_{\mbox{\tiny $\begin{array}{c}n_1,n_2\geq 1 \\n_1\equiv n_2 - \alpha(\bm) \ \mbox{mod}\ p\end{array}$}}[-n_1-n_2+\ii\alpha(\sigma_0)]}\right]~.
\eea
We then rewrite this as a product over only the positive roots $\alpha>0$, while at the same time multiplying by the 
same expression with  $\alpha\rightarrow -\alpha$. In doing this, 
one sees that all the terms in the numerators and denominators cancel, \emph{except}
for the numerator contributions of $\{n_1=0,n_2\geq 1\}$, $\{n_1\geq 1,n_2=0\}$, which are left uncancelled. We are thus left with
\bea
Z^{\mathrm{vector}}_{\mathrm{1-loop}}(\sigma_0,\bm) & =& \prod_{\alpha>0} \Bigg[\prod_{r\geq 1}\left[-\alpha(\bm)+pr+\ii\alpha(\sigma_0)\right]
\frac{\prod_{r\geq 0}\left[\alpha(\bm)+pr-\ii\alpha(\sigma_0)\right]}{(-\ii\alpha(\sigma_0))^{\delta_{\alpha(\bm),0}}}\times\nonumber\\
&& 
\frac{\prod_{r\geq 0}\left[\alpha(\bm)+pr+\ii\alpha(\sigma_0)\right]}{(\ii\alpha(\sigma_0))^{\delta_{\alpha(\bm),0}}}\prod_{r\geq 1}\left[-\alpha(\bm)+pr-\ii\alpha(\sigma_0)\right]\Bigg]~.
\eea
Notice that here, in a slight abuse of notation, we have assumed that $0\leq \alpha(\bm)<p$.
The last equation may then be rewritten
\bea
Z^{\mathrm{vector}}_{\mathrm{1-loop}}(\sigma_0,\bm) & =& \prod_{\alpha>0}\left[\prod_{r=1}^\infty (pr)^4\right]
 (\alpha(\bm)-\ii\alpha(\sigma_0))\prod_{r=1}^\infty\left[1+\frac{\left(\alpha(\sigma_0)+\ii\alpha(\bm)\right)^2}{r^2p^2}\right]\times\nonumber\\
&&(\alpha(\bm)+\ii\alpha(\sigma_0)) \prod_{r=1}^\infty \left[1+\frac{\left(\alpha(\sigma_0)
-\ii\alpha(\bm)\right)^2}{r^2p^2}\right]\cdot\frac{1}{(\alpha(\sigma_0)^2)^{\delta_{\alpha(\bm),0}}}\nonumber
\eea
Zeta function regularizing $\prod_{r=1}^\infty (pr)^4 \stackrel{\mathrm{zeta}}{=} (2\pi)^2/p^2$ and using the infinite product formula 
for $\sinh$, we obtain
 \cite{Gang:2009wy, Benini:2011nc}
\bea\label{vector}
Z^{\mathrm{vector}}_{\mathrm{1-loop}}(\sigma_0,\bm) \ =\  \prod_{\alpha>0} \frac{2\sinh\left[\frac{\pi}{p}\left(\alpha(\sigma_0)+\ii\alpha(\bm)\right)\right]2\sinh\left[\frac{\pi}{p}\left(\alpha(\sigma_0)-\ii\alpha(\bm)\right)\right]}{\left(\alpha(\sigma_0)^2\right)^{\delta_{\alpha(\bm)},0}}~.
\eea

\subsection{Partition function}

Putting everything together from the previous sections, we arrive at the final formula 
for the localized partition function of an $\mathcal{N}=2$ Chern-Simons-matter
 theory on the Lens space $L(p,1)$
\bea
\label{partitionfunction} Z \ =\  \sum_{\bm}\int \diff \sigma_0 \exp\left[\frac{\ii \pi k}{p}\left(\mathrm{Tr} (\sigma_0^2)-\mathrm{Tr}\, (\bm^2)\right)\right]
Z^{\mathrm{vector}}_{\mathrm{1-loop}}(\sigma_0,\bm) \,  Z^{\mathrm{matter}}_{\mathrm{1-loop}}(\sigma_0,\bm)~,
\eea
where the one-loop vector and matter contributions are given by (\ref{vector}), (\ref{matter}), respectively.

Recall that for a $U(N)$ gauge group, $\sigma_0$ is a  constant $N\times N$ Hermitian matrix
that commutes with $\bm$. We may thus diagonalize
\bea
\sigma_0 &=& \left(\frac{\lambda_1}{2\pi},\ldots,\frac{\lambda_N}{2\pi}\right)~,
\eea 
where $\lambda_i/2\pi$, $i=1,\ldots,N$, are the real eigenvalues of $\sigma_0$, and we order $\lambda_1\leq \cdots\leq\lambda_N$. A choice of positive roots 
for $G$ is then
\bea
\alpha_{ij}(\sigma_0) &=& \frac{\lambda_i-\lambda_j}{2\pi}~,
\eea
with $i<j$. Notice that the Vandermonde determinant then contributes a factor 
to the integrand  of (\ref{partitionfunction}) 
when rewriting
\bea
\int \diff\sigma_0 &=& \int \prod_{i=1}^N\frac{\diff\lambda_i}{2\pi} \prod_{i<j \, |\, m_i=m_j}\left(\frac{\lambda_i-\lambda_j}{2\pi}\right)^2~,
\eea
 which precisely cancels 
the denominator in (\ref{vector}).

We now specialise to the particular gauge theory of interest, namely the ABJM theory on $N$ M2-branes 
\cite{Aharony:2008ug}. This is a $U(N)\times U(N)$ Chern-Simons gauge theory with 
Chern-Simons levels $(k,-k)$ for the two gauge group factors,
two chiral matter fields in the bifundamental representation $(\mathbf{N},\overline{\mathbf{N}})$, 
and two chiral matter fields in the conjugate  $(\overline{\mathbf{N}},\mathbf{N})$ representation.
More precisely, this is the low energy worldvolume theory on $N$ M2-branes 
transverse to $\C^4/\Z_k$, where the $\Z_k$ acts with weights $(1,1,-1,-1)$ on the four complex coordinates.  
The R-charges/scaling dimensions of the 4 chiral fields all take the canonical value of 
$\Delta=\frac{1}{2}$. We may thus introduce eigenvalues 
$\lambda_i$, $\tilde\lambda_i$, $i=1,\ldots,N$, for the two 
gauge group factors, and correspondingly matrices
$\bm$, $\tbm$ specifying the flat connections 
for each copy of $U(N)$. Note that the weights 
for the bifundamental representation $(\mathbf{N},\overline{\mathbf{N}})$
are 
\bea
\rho_{ij}(\sigma_0) &=& \frac{\lambda_i-\tilde{\lambda}_j}{2\pi}~,
\eea
with minus this for the conjugate representation. Thus the partition function 
for the ABJM theory on $L(p,1)$ is
\bea\label{ABJM}
Z &  =&  \sum_{\bm,\tilde{\bm}}\frac{1}{N!^2}\int \prod_{i=1}^N \frac{\diff\lambda_i}{2\pi}\frac{\diff\tilde\lambda_i}{2\pi}
\exp\left[\frac{\ii k}{4\pi p}\left(\sum_{i=1}^N \left( \lambda_i^2-\tilde\lambda_i^2\right)-(2\pi)^2\sum_{i=1}^N 
\left(m_i^2-\tilde{m_i}^2\right)\right)\right]\times\nonumber\\
 &&\prod_{i<j}2 \sinh\left[\frac{\lambda_i-\lambda_j+2\pi\ii (m_i-m_j)}{2p}\right]2\sinh\left[\frac{\lambda_i-\lambda_j-2\pi\ii(m_i-m_j)}{2p}\right]\times\nonumber\\
&& \prod_{i<j}2 \sinh\left[\frac{\tilde\lambda_i-\tilde\lambda_j+2\pi\ii (\tilde{m}_i-\tilde{m}_j)}{2p}\right]2\sinh\left[\frac{\tilde\lambda_i-\tilde\lambda_j-2\pi\ii(\tilde{m}_i-\tilde{m}_j)}{2p}\right]\times\nonumber\\
&& \prod_{i,j}\left[P_{p}^{m_i-\tilde{m}_j}\bigg(\frac{\lambda_i-\tilde\lambda_j}{2\pi}\bigg)\right]^2~, 
\eea
where we have defined
\bea\label{P}
P_{p}^\kappa(\al) &:= & \prod_{\ell=0}^\infty \left(\frac{\ell+\frac{3}{2}+\ii \al}{\ell+\frac{1}{2}-\ii \al}\right)^{N_\kappa(\ell)}
\left(\frac{\ell+\frac{3}{2}-\ii \al}{\ell+\frac{1}{2}+\ii \al}\right)^{N_{p-\kappa}(\ell)}~.
\eea
The latter is precisely the contribution from \emph{one}  $(\mathbf{N},\overline{\mathbf{N}})$ 
chiral field, and \emph{one}  $(\overline{\mathbf{N}},\mathbf{N})$ field, and 
these correspond to the respective factors in the product (\ref{P}). 
As before, the notation $N_\kappa(\ell)$ means the number of 
half-integers $n\in\{-\frac{\ell}{2},-\frac{\ell}{2}+1,\ldots,\frac{\ell}{2}\}$ 
such that $2n\equiv \kappa$ mod $p$. 
The square at the very end of (\ref{ABJM}) then 
accounts for the fact there are two of each type of 
bifundamental field. 

\subsection{Matter potentials}

The matter potentials
\bea\label{V}
V_{p}^\kappa(\al) &:=  & \log P_{p}^\kappa(\al)~,
\eea
where $P_{p}^\kappa(\al)$ is defined by (\ref{P}), play an important role in the dynamics 
of the matrix model (\ref{ABJM}). A general discussion of these potentials, 
which in general involve polygamma functions, may be found in appendix \ref{app:potentials}. 
In particular, the products in (\ref{P}) are divergent and must be regularized, and we 
do this using zeta function regularization.
The resulting functions simplify somewhat in particular cases. In this subsection we give a few examples, for low values of $p$.

Recall that for fixed $p\geq 1$, we have $0\leq \kappa<p$. 
The (regularized) matter potentials for small $p$ then simplify to
\bea
 V_{1}^{0}(\al) & = & -\log \left[2\cosh(\pi\al)\right]~,\nonumber\\
V_{2}^0(\al) &=& -\frac{1}{2} \log [2 \cosh(\al \pi)] + 2\al \cot^{-1} \ex^{\pi \al} +\frac{\ii}{\pi} \left[\mathrm{Li}_2(-\ii \ex^{-\al \pi}) - \mathrm{Li}_2(\ii \ex^{-\al \pi}) \right]~,\nonumber \\
V_{2}^1(\al) &=& -\frac{1}{2} \log (2 \cosh(\al \pi) ) -2\al \cot^{-1} \ex^{\pi \al} -\frac{\ii}{\pi} \left[ \mathrm{Li}_2(-\ii \ex^{-\al \pi}) - \mathrm{Li}_2(\ii \ex^{-\al \pi}) \right]~,\nonumber \\
V_{3}^0(\al) &=& \log \frac{2 \cosh^2(\pi \al/3)}{\cosh \pi \al}~,\nonumber\\
V_{3}^1(\al) &=&V_{3,1}^2(\al) \ = \ -\log [2 \cosh (\pi \al/3)]~.
\eea
The reader is referred to the Appendices \ref{app:potentials} and \ref{app:asymptotics} for more
detail.


\section{The large $N$ limit: numerical results}\label{sec:numerics}

Our aim in the remainder of the paper is to compute the M-theory limit of the ABJM partition function (\ref{ABJM}), 
which means fixed Chern-Simons level $k$ and  $N\rightarrow\infty$. 
The partition function (\ref{ABJM})  is an extremely complicated object, and to gain some intuition we will begin 
in this section by solving the matrix model numerically for large values of $N$. We will do this 
by extending the saddle-point methods of \cite{Herzog:2010hf} to the present case. 
The behaviour is simple enough to suggest an 
ansatz for the eigenvalue distribution, precisely as for the ABJM model on $S^3$ studied in \cite{Herzog:2010hf}, which in section \ref{sec:analytic} we then analytically show 
reproduces the numerics. Moreover, this analytic result agrees with 
 the expected large $N$ gravity dual result for the free 
energy of $F_{\mathrm{M}-\mathrm{theory}} = \frac{\pi\sqrt{2k}}{3p}N^{3/2}$. 

\subsection{General discussion}

The matrix model partition function (\ref{ABJM}) of $N$ M2-branes on a  Lens space $L(p,1)$ has the following form
\bea\label{Zm}
Z & = & \sum_{\bm,{\tilde\bm}} Z_{\bm,\tbm} \ = \ \sum_{\bm,\tbm} \int \left( \prod_{i=1}^N \dd \lambda_i \dd\tilde \lambda_i  \right) \ex^{-F_{\bm,\tbm }(\lambda_i,\tilde \lambda_i) }~,
\eea
where recall that $\bm=\mathrm{diag}(m_1,\ldots,m_N)$ has entries $m_1\leq m_2\leq \ldots\leq m_N$ with $0\leq m_i<p$, and specifies 
the flat connection for the first $U(N)$ gauge group, while tilded quantities refer to the second $U(N)$ gauge group.
The basic idea is that when the number $N$ of eigenvalues $\lambda_i$ is large, each contribution $Z_{\bm,\tbm}$  can be well approximated in the saddle-point limit by $Z=\ex^{-F}$, where the free energy $F$ is an extremum of $F_{{\bf m},{\bf \tilde m}}(\lambda_i,\tilde \lambda_i)$ with respect to $\lambda_i$ and $\tilde \lambda_i$.
 
 Given $F(\lambda_i,\tilde \lambda_i)$ (in what follows, we suppress the lower indices ${\bf m,\tilde m}$) the saddle-point equations are
 
 \begin{equation}
 \label{saddle}
\frac{\partial F}{\partial \lambda_i} \  = \ 0~,~~~~~\frac{\partial F}{\partial \tilde \lambda_i} \ = \ 0~.
 \end{equation}
The extremum of the free energy is then given by $F(\lambda^{0}_i,\tilde \lambda^{0}_i)$, where $\lambda^{0}_i,\tilde \lambda^{0}_i$ are the solutions of the saddle-point equations. This then gives the leading contribution to $Z_{\bf m,\tilde m}$ at large $N$ and fixed $k$. As 
for $p=1$, it will turn out that the saddle-point solution has \emph{complex} eigenvalues, which means 
we deform the real integrals over $\lambda_i,$ $\tilde\lambda_i$ in (\ref{Zm}) into the complex plane.

Even though highly non-trivial, the equations (\ref{saddle}) can be solved numerically. It is convenient to view these equations as describing the equilibrium configuration of $2N$ particles, whose two-dimensional coordinates are given by the complex numbers $\lambda_i$ and $\tilde \lambda_i$. This equilibrium configuration can be found by introducing a ``time dependence'', so that $\lambda_i,~\tilde \lambda_i \rightarrow \lambda_i(t),~\tilde \lambda_i(t)$, and writing down equations of motion for $\lambda_i(t)$ and $\tilde \lambda_i(t)$ such that their solutions approach the equilibrium configuration for late times:\footnote{In order for the eigenvalues to go to the correct attractor point as $t \rightarrow \infty$, {\it a priori} one might need to multiply the left hand sides of (\ref{saddletime}) by a complex number. For the case at hand in fact this is unnecessary.}
 \begin{equation}
 \label{saddletime}
\frac{\dd \lambda_i}{\dd t}\ = \ -\frac{\partial F}{\partial \lambda_i}~,~~~~~\frac{\dd \tilde\lambda_i}{\dd t}\  = \ - \frac{\partial F}{\partial \tilde \lambda_i}~.
 \end{equation}
In the following we will solve these equations numerically. From this we will extract generic behaviour that will lead to a corresponding analytic computation in section \ref{sec:analytic}.
 
 \subsection{Flat connection dependence}
 
 A new ingredient in the partition function on Lens spaces, with respect to the case on $S^3$, is the sum over different flat connections  labelled by $\bm$, $\tbm$. We are interested in the large $N$ limit, and in this limit we expect
  \begin{equation}\label{Zlimit}
 Z\ =\  \sum_{{\bf m},{\bf \tilde m}} \ex^{-F_{{\bf m},{\bf \tilde m}}} \ \longrightarrow \  \ex^{-F_{\mathrm{M-theory}}} \ = \ Z_{\mathrm{M-theory}} ~.
 \end{equation}
In the supergravity approximation to M-theory, we are computing the  log of the partition function in the large $N$ limit, 
and we are interested in the leading term only. Hence, even though we have the sum of many terms on the left hand side of (\ref{Zlimit}), we expect only certain terms to contribute. More precisely, we may focus on the contribution of $(\bm,\tbm)=(\bm_0,\tbm_0)$ with least $F_{{\bf m}_0,{\bf \tilde m}_0}$, in the large $N$ limit. Note that contributions of $\ex^{-F_{{\bf m},{\bf \tilde m}}}$ for other choices of $(\bm,\tbm)$ do not  need to be suppressed with respect to that for $(\bm_0,\tbm_0)$: if they give a similar contribution, this will simply lead to a logarithmic, and hence subleading, correction at large $N$, since we are taking the logarithm to obtain the free energy.
 
We have performed a numerical evaluation of $F_{{\bf m},{\bf \tilde m}}$, for several values of $N,p$ and all choices of ${\bf m}, {\bf \tilde m}$. In all  cases the leading contribution comes from ${\bf m} ={\bf \tilde m} = \mathrm{diag}(c,c,c,...,c) = c \cdot 1_{N\times N}$, where $c$ is an integer with $0\leq c<p$.
  Hence the first lesson we draw from the numerical analysis is that we may focus on the specific case  ${\bf m} ={\bf \tilde m} = \mathrm{diag}(c,c,c,...,c)$ if we are only interested in the large $N$ limit. This  simplifies the problem, and its treatment, enormously. Furthermore, note that with these choices of 
flat connection the eigenvalues will respect certain symmetries, discussed below, but that this would not be true for any other choice 
of flat connection. So from now on we focus on this case.\footnote{Other choices of $\bm$, $\tbm$, presumably important for computing 
subleading corrections, are briefly discussed in Appendix B.}
 
\subsection{Numerical plots} 

Figure \ref{fig:p2} shows the distribution of eigenvalues for the case $p=2$ and $N=100$. 
From this distribution we can draw several conclusions. First, we see that the eigenvalue distribution is invariant under $\lambda_i \rightarrow - \lambda_i$, $\tilde \lambda_i \rightarrow - \tilde \lambda_i$. Furthermore, for the equilibrium configuration $\lambda_i$ and $\tilde \lambda_i$ are complex conjugates of each other. To be more precise, we find that $\lambda_{i} = - \lambda_{N-i+1}$ (with the same for $\tilde \lambda_i$) and $\bar\lambda_i = {\tilde \lambda}_i$. As for the $p=1$ case, these are symmetries of the equations of motion, so we expect these symmetries for the 
equilibrium distributions as well. As already mentioned, however, these symmetries will not be present for more general choices of ${\bf m},{\bf \tilde m} $.
 
 \begin{figure}[ht!]
\centering
\includegraphics[scale=1]{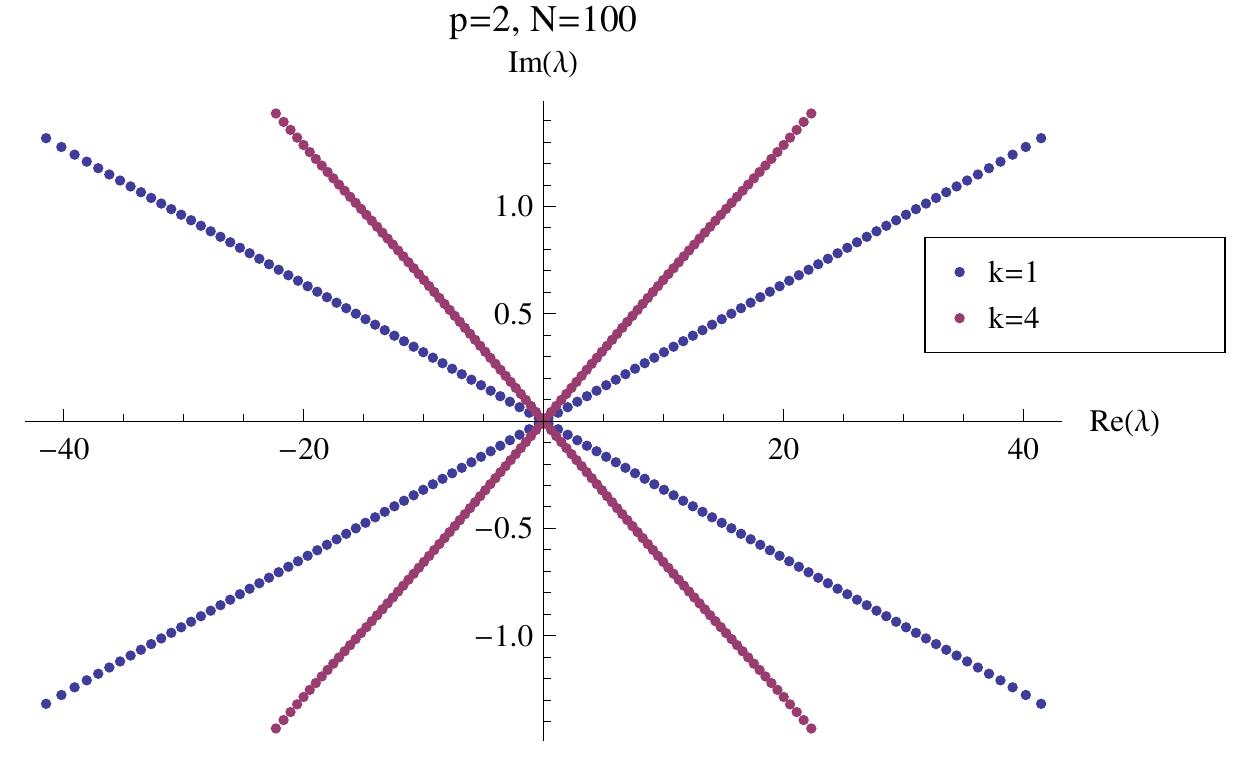}
\caption{Distribution of eigenvalues for $p=2$, $N=100$.}\label{fig:p2}
\end{figure}
 
\begin{figure}[ht!]
\centering
\includegraphics[scale=1]{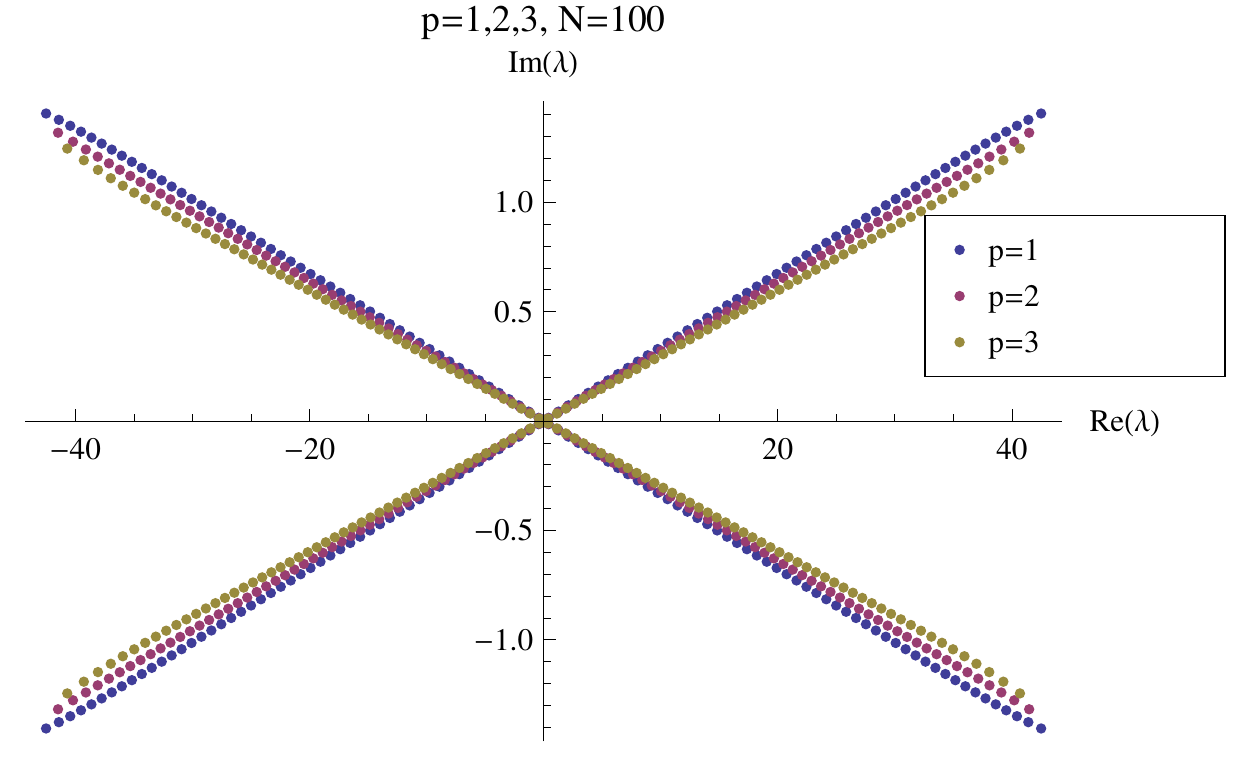}
\caption{Distribution of eigenvalues for $p=1,2,3$ and $N=100$.}\label{fig:p123}
\end{figure}

Other features are that for large values of $N$ the density of eigenvalues is relatively uniform, the real part of the eigenvalues grows with $N$, while the imaginary part stays bounded. The numerics are consistent with the real part growing as $N^{1/2}$, while the imaginary part of the eigenvalues stays bounded between $-\pi/2$ and $\pi/2$. 
 As we increase the Chern-Simons level $k$ the slope also increases. The analytic treatment in section \ref{sec:analytic} (after assumptions justified by the numerics)  predicts a slope proportional to $\sqrt{k}$ -- in fact precisely the same slope as for $p=1$.  This is also consistent with the numerical 
results -- see Figure \ref{fig:p123}.

\section{The large $N$ limit: analytic results}\label{sec:analytic}

As in the previous section, the idea is to compute the partition function (\ref{ABJM}) in a saddle-point approximation, 
focusing on the contribution from $\bm=\tbm=c\cdot 1_{N\times N}$, which from the numerics 
we see determines the free energy in the large $N$ limit.
As the number of eigenvalues
$N$ for each gauge group tends to infinity, one has a continuum limit in which one can
replace the sums over eigenvalues in the potential by integrals. In particular, one can then
separate the interactions between eigenvalues into ``long range forces,'' for which the
interaction between eigenvalues is non-local, plus a local interaction. A key point 
is that
 these long range forces automatically cancel. An appropriate ansatz 
for $\lambda_i$ will then lead to a simple local action for the eigenvalues, 
which may be solved in the saddle-point approximation exactly 
in the large $N$ limit. This analytic result may then be checked 
against the numerical results, and we find excellent agreement. 
We will also comment on the relation to the gravity dual.

\subsection{Long range forces}

Let us focus first on the long range forces, which come 
from the leading terms in an asymptotic expansion of 
the $\sinh$ and matter potential $V_{p}^{\kappa=0}(\al)$ in 
(\ref{ABJM}).\footnote{Recall here that since $\bm=\tbm$ is proportional to the identity matrix, 
we have $\kappa=0$ in all cases.} In the former case we define
\bea\label{sinhlong}
\left[\log 2\sinh \al\right]^{\as} &:= & \al \, \mathrm{sign}\left(\mathrm{Re}\, \al\right)~.
\eea
The point here is that for $\mathrm{Re}\, \al>0$ we have the series
\bea\label{sinhplus}
\log \left[2\sinh \al\right] &=& {z} - \sum_{\ell=1}^\infty \frac{1}{\ell}\ex^{-2\ell\al}~.
\eea
while for $\mathrm{Re}\, \al<0$ we have
\bea\label{sinhminus}
\log \left[2\sinh \al\right] &=& \ii\pi -{z} -\sum_{\ell=1}^\infty \frac{1}{\ell}\ex^{2\ell\al}~.
\eea
We shall see momentarily that the constant $\ii\pi$ term in (\ref{sinhminus}) does not contribute to the long range 
force computation, which is why we omit this constant in the definition (\ref{sinhlong}). 
The sums of exponential terms in (\ref{sinhplus}), (\ref{sinhminus})  will be of relevance momentarily.

The matter potentials $V_{p}^0(\al)$ depend in a complicated way on $p$.
The relevant asymptotic expansions are discussed in appendix \ref{app:asymptotics}. In particular, 
this leads to
\bea\label{Vlong}
\left[V_{p}^0(\al)\right]^\as &:= & -\frac{\pi \al}{p}\, \mathrm{sign}\left(\mathrm{Re}\, \al\right)~.
\eea
We then have the following general form of the expansions for $V_{p}^0(\al)$
\bea\label{Vexpansion}
V_{p}^0(\al) \, = \, \left[V_{p}^0(\al)\right]^\as  + \pi\al\, \mathrm{sign}(\mathrm{Re}\, z)\sum_{\ell\in \frac{1}{p}\mathbb{N}} c_\ell\, \ex^{-\ell\pi\al\, \mathrm{sign}(\mathrm{Re}\, z)}  + \sum_{\ell\in \frac{1}{p}\mathbb{N}} d_\ell\, \ex^{-\ell\pi\al\,  \mathrm{sign}(\mathrm{Re}\, z)} ~,
\eea
for appropriate constants $c_\ell$, $d_\ell$ depending on $p$.

In the large $N$ limit we then take a continuous limit, in which sums over $i=1,\ldots,N$ become Riemann integrals
\bea\label{cont}
\frac{1}{N}\sum_{i=1}^N & \longrightarrow & \int_{x_{\mathrm{min}}}^{x_{\mathrm{max}}} \rho(x)\, \diff x.
\eea
The numerical results of section \ref{sec:numerics} then  suggest we make the following ansatz for the eigenvalues
\bea\label{lambdax}
\lambda(x) &=& N^\alpha x + \ii y(x)~, \qquad \tilde{\lambda}(x) \ = \ N^\alpha x  - \ii y(x)~,
\eea
where $\alpha>0$, and these formulae are understood to be correct to order $N^{-\epsilon}$, for some $\epsilon>0$.
Notice we have deformed the real eigenvalues of the Hermitian matrix $\sigma$ into the complex plane in (\ref{lambdax}), 
anticipating a complex saddle point, and that the function $\rho(x)$ describes the density of the eigenvalues. 
Also recall that $\bar\lambda_i\leftrightarrow\tilde\lambda_i$ is a symmetry of the system -- in 
(\ref{lambdax}) we have imposed that the solution is invariant under this symmetry, which is 
again supported by the numerical results.

The long range forces are then, by definition, determined by the leading asymptotic terms in the potential. 
Substituting (\ref{sinhlong}),  (\ref{Vlong}) into the logarithim of the partition function (\ref{ABJM}) and taking the
continuum limit (\ref{cont}), we obtain
\bea\label{Flong}
-F_{\as} &=& N^2 \int_{x_{\mathrm{min}}}^{x_{\mathrm{max}}} \rho(x)\, \diff x \int_{x_{\mathrm{min}}}^{x_{\mathrm{max}}} \rho(x')\, \diff x' 
\, \mathrm{sign}(x-x')\Bigg[\frac{1}{2p}(\lambda(x)-\lambda(x')) \nonumber\\
&& + \frac{1}{2p}(\tilde\lambda(x)-\tilde\lambda(x)') - 2\times \frac{\lambda(x)-\tilde\lambda(x')}{2p}\Bigg]~.
\eea
Here one sees from (\ref{ABJM}) that one substitutes $z=(\lambda(x)-\tilde\lambda(x'))/2\pi$ into 
(\ref{Vlong}). The factor of 2 in the last term of (\ref{Flong}) accounts for the two copies 
of chiral fields in the $(\mathbf{N},\overline{\mathbf{N}})$ and  $(\overline{\mathbf{N}},\mathbf{N})$ representations.
Also notice that the original sum in the vector multiplet contribution to (\ref{ABJM}) is over $i<j$, which means
$x-x'<0$ in the continuum limit. In writing (\ref{Flong}) we have simply extended this to a sum over $i>j$ by 
replacing $\lambda_i-\lambda_j$ by $\lambda_j-\lambda_i$. 
It is then straightforward to see from the ansatz (\ref{lambdax}) that all the terms in (\ref{Flong}) cancel: 
the real parts simply cancel inside the square bracket, while the imaginary parts contribute zero 
on using the anti-symmetry under $x\leftrightarrow x'$ implied by the $\mathrm{sign}(x-x')$ term. 
Thus $F_\as=0$, and the long range forces indeed cancel for $L(p,1)$.

\subsection{Local action}

It follows that only the exponential sums in (\ref{sinhplus}), (\ref{sinhminus}), (\ref{Vexpansion})
contribute to the partition function in the large $N$ limit. Since the latter depend on $\mathrm{sign}\left(\mathrm{Re}\, z\right)$, which is
equal to $\mathrm{sign}(x-x')$, we first split the double integrals as
\bea
N^2 \int_{x_{\mathrm{min}}}^{x_{\mathrm{max}}} \rho(x)\, \diff x \int_{x_{\mathrm{min}}}^{x_{\mathrm{max}}} \rho(x')\, \diff x' 
&\longrightarrow & N^2 \int_{x_{\mathrm{min}}}^{x_{\mathrm{max}}} \rho(x)\, \diff x \int_{x_{\mathrm{min}}}^{x} \rho(x')\, \diff x' 
\nonumber\\
&& + N^2 \int_{x_{\mathrm{min}}}^{x_{\mathrm{max}}} \rho(x)\, \diff x \int_{x}^{x_{\mathrm{max}}} \rho(x')\, \diff x'~,
\eea
so that $x-x'>0$ for the first term on the right hand side, while $x-x'<0$ for the second term. 
We will then apply the general formula
\bea\label{delta}
\int_{x_\mathrm{min}}^x\diff x' \, \ex^{-\beta N^\alpha(x-x')}\, f(x,x') &=& \frac{1}{\beta N^\alpha}\left[\ex^{-\beta N^\alpha(x-x')}\, f(x,x')\right]_{x_{\mathrm{min}}}^x\nonumber\\
&& - \frac{1}{\beta N^\alpha}\int_{x_{\mathrm{min}}}^x \diff x'\, \ex^{-\beta N^\alpha(x-x')}\, \frac{\diff}{\diff x'} f(x,x')~,
\eea
which follows trivially from an integration by parts. The first term on the right hand side is simply $\frac{1}{\beta N^{\alpha}}f(x,x)$, 
plus a term which is exponentially suppressed in the large $N$ limit. The formula (\ref{delta}), with a similar formula applying for $x-x'<0$,  amount to the representation
\bea
\delta(x) &=& \lim_{c\rightarrow\infty} \frac{c}{2}\, \ex^{-c|x|}~,
\eea
thus
reducing the integral over $x$, $x'$ to an integral over $x$, in the large $N$ limit.
Applying this to the sums over exponentials in (\ref{sinhplus}), (\ref{sinhminus}), (\ref{Vexpansion}) is a straightforward 
task. For the vector multiplet and matter multiplet contributions, we obtain to leading order
\bea\label{Fvector}
-F_{\mathrm{vector}}&=& -\frac{4p\pi^2}{6}N^{2-\alpha}\int_{\xmin}^{\xmax} \rho(x)^2\, \diff x +o(N^{2-\alpha})~,\\
\label{Fmattersum}
-F_{\mathrm{matter}} &=& 8N^{2-\alpha}\int_{\xmin}^{\xmax} \rho(x)^2\, \diff x \Bigg\{\sum_{\ell\in \frac{1}{p}\mathbb{N}}
\frac{c_\ell}{\ell}y(x)\sin \left[\ell y(x)\right]+ \frac{c_\ell}{\ell^2}\cos \left[\ell y(x)\right] \nonumber\\
&&
+ \frac{d_\ell}{\ell}\cos \left[\ell y(x)\right]\Bigg\}+o(N^{2-\alpha})~.
\eea
The term in curly brackets is denoted $J_{p}[y(x)]$ in Appendix \ref{app:asymptotics}, and may be evaluated 
by Fourier summation to give
\bea
-F_{\mathrm{matter}}&=& 8N^{2-\alpha}\int_{\xmin}^{\xmax} \rho(x)^2\, \diff x\left[\frac{\pi^2}{24}\left(2p-\frac{3}{p}\right) + \frac{y(x)^2}{2p}\right] +o(N^{2-\alpha})~.
\eea
Combining with (\ref{Fvector}), we thus obtain the leading order result
\bea
F_{\mathrm{one-loop}} \ = \ F_{\mathrm{vector}} + F_{\mathrm{matter}} \, = \,  \frac{N^{2-\alpha}}{p}\int_{\xmin}^{\xmax} \rho(x)^2\, \left[\pi^2 - 4y(x)^2\right]\, \diff x+o(N^{2-\alpha})~.
\eea
It remains the add the contribution of the classical terms in (\ref{ABJM}). This is a trivial modification of the $p=1$ computation, the only 
difference being the factor of $1/p=\mathrm{Vol}(L(p,1))/\mathrm{Vol}(S^3)$ :
\bea
F_{\mathrm{classical}} &=& \frac{kN^{1+\alpha}}{p\pi}\int_{\xmin}^{\xmax} xy(x)\rho(x)\, \diff x + o(N^{1+\alpha})~.
\eea
The total free energy action is then to leading order
\bea\label{Faction}
F &=&  \frac{kN^{1+\alpha}}{p\pi}\int_{\xmin}^{\xmax} xy(x)\rho(x)\, \diff x + \frac{N^{2-\alpha}}{p}\int_{\xmin}^{\xmax} \rho(x)^2\, \left[\pi^2-4y(x)^2\right] \, \diff x~.
\eea
As for the case of the round sphere with $p=1$, non-trivial saddle points will require both terms to be of the same 
order, so that $\alpha=\frac{1}{2}$ and hence $\lambda(x)=N^{1/2}x+\ii y(x)$.

Remarkably, we see that 
the action $F$ in (\ref{Faction}) is simply $1/p$ times the action for $p=1$ in reference \cite{Herzog:2010hf}. In particular, the saddle point equations derived from 
(\ref{Faction}) are identical to those in reference \cite{Herzog:2010hf}, which allows us to simply write down
that the density $\rho(x)$ is constant
\bea
\rho(x) &=& \frac{k}{2\sqrt{2}\pi}~,
\eea
and the imaginary part of the eigenvalues $y(x)$ is linear
\bea
y(x) &=& \frac{\sqrt{k}}{2\sqrt{2}}\, x~,
\eea
with $-\xmin=\xmin = \pi\sqrt{2/k}$, so that $y(x)\in\left[-\frac{\pi}{2},\frac{\pi}{2}\right]$. Of course, 
this is perfectly consistent with the numerical results in Figures \ref{fig:p2} and \ref{fig:p123}.
The dependence on $p$ only enters in the free energy $F$ evaluated on this saddle point solution, 
which is
\bea\label{finalfree}
F &=& N^{3/2}\frac{\pi\sqrt{2k}}{3p} +o(N^{3/2})~.
\eea
Again, this is consistent with the numerics. 

The formula (\ref{finalfree}) is expected from the supergravity dual solution
AdS$_4/\Z_p\times S^7/\Z_k$, since the quotient by $\Z_p$ simply divides the 
overall supergravity action by $p$. The only slight subtlety here is that 
AdS$_4/\Z_p$ has a $\Z_p$ orbifold singularity at the ``centre''. In principle 
there might exist degrees of freedom at this singularity which then contribute 
to the leading order large $N$ free energy, but the field theory result we have obtained implies 
this is not the case.

\section{Outlook}\label{sec:outlook}

In this paper we considered the large $N$ limit of the partition function of $N$ M2-branes 
on the Lens space $L(p,1)$. Some open problems include:
\begin{itemize}
\item The partition function (\ref{ABJM}) is valid for all $N, k, p$, 
and it would be interesting to study this more generally, for example at finite $N$, or in the 't Hooft 
limit in which $N/k$ is held fixed. 
\item One might also consider squashed Lens spaces, for which there 
are supergravity dual solutions \cite{Martelli:2011fw}.
\item Another interesting open question is whether these theories have a description 
in terms of a Fermi gas, as for the ABJM theory on $S^3$ \cite{Marino:2011eh}. This 
may be a useful method for computing subleading corrections.
\end{itemize}
The generalization of these results to more general Lens space $L(p,q)$ will be addressed in a forthcoming publication.

\subsection*{Acknowledgments}
\noindent
L. F. A and J. F. S. would like to thank the Isaac Newton Institute for hospitality 
during the completion of this work.
 J.~F.~S. is supported by a Royal Society Research Fellowship. 

\appendix

\section{Computation of  potentials}\label{app:potentials}
  
In this appendix we present analytical expressions for the infinite products that enter into the partition function for the Lens spaces $L(p,1)$. These infinite products have the following form:  
\begin{equation}
P_{p}^\kappa(z) \ = \ \prod_{\ell=0}^\infty \left(\frac{\ell+\tfrac{3}{2}+\ii \al }{\ell+\tfrac{1}{2}-\ii \al} \right)^{N_\kappa (\ell)}  \left(\frac{\ell+\tfrac{3}{2}-\ii \al }{\ell+\tfrac{1}{2}+\ii \al} \right)^{N_{p-\kappa}(\ell)}~,
\end{equation}
where $\kappa=0,1,...,p-1$ and $N_\kappa(\ell)$ denotes the number of integers $m=\{-\ell,-\ell+2,...,\ell-2,\ell \}$ such that $m\equiv \kappa$ mod $p$. For computing the free energy the log of this product is relevant. Hence we introduce the potentials $V_{p}^\kappa$:
\begin{equation}
V_{p}^\kappa \ :=\  \log P_{p}^\kappa~.
\end{equation}

\subsection{$p=1$} 

Let us explain in detail how to obtain the potential for $p=1$. This result is already known \cite{Kapustin:2009kz}, but it is instructive to recover it. In this case we have $\kappa=0$ and the potential reduces to
\begin{equation}
V_1(\al) \  = \ \sum_{\ell=0}^\infty   (\ell+1) \log \left(\frac{\ell+\tfrac{3}{2}+\ii \al }{\ell+\tfrac{1}{2}-\ii \al}\cdot  \frac{\ell+\tfrac{3}{2}-\ii \al }{\ell+\tfrac{1}{2}+\ii \al} \right)~.
\end{equation}
This sum is divergent, hence in order to compute it we need to regularize it. A standard procedure is to take the derivative of the potential and perform the sum. We obtain
\begin{equation}
V'_1(\al) \ = \ -\pi \tanh\,  (\pi \al)~.
\end{equation}
Of course, in taking the derivative we are dropping an additive constant (which could be infinite). Integrating back we find
\begin{equation}
V_1(\al) \ =\  -\log \cosh \, (\pi \al) +c ~.
\end{equation}
The integration constant can be fixed by zeta function regularization. This is explained in detail for instance in appendix A of \cite{Drukker:2010nc}. The constant $c$ is defined as the value of $V_1(\al)$ at $\al=0$. We obtain
\begin{equation}
\frac{1}{2}c \ = \ \sum_{\ell=0}^\infty   (\ell+1) \log \left(\frac{\ell+\tfrac{3}{2} }{\ell+\tfrac{1}{2}} \right)~.
\end{equation}
We will compute this divergent sum by using $\zeta-$function regularization. Let us define
\bea
\zeta_Z(s) &=& \sum_{\ell=0}^\infty \left( \frac{\ell+1}{(\ell+\tfrac{3}{2})^s}- \frac{\ell+1}{(\ell+\tfrac{1}{2})^s} \right)~.\eea
Hence the quantity we wish to compute is just $-\zeta'_Z(0)$. These sums are by definition zeta functions, and their generalization, Hurwitz zeta functions
\begin{equation}
\zeta_a(s) \ = \ \sum_{\ell=0}^\infty \frac{1}{(\ell+a)^s}~.
\end{equation}
Sums with factors of $\ell$ in the numerator are easily obtained, since a factor of $\ell+a$ in the numerator can be absorbed by a shift $s \rightarrow s-1$. For the particular case at hand we obtain
\bea
\zeta_Z(s)  &=& -(2^s - 1) \zeta(s)~.\eea
Hence $-\zeta'_Z(0)= -\frac{\log 2}{2}$. This implies $c=-\log 2$, and the following result for $V_1$:
\begin{equation}
V_1(\al) \ = \ -\log (2 \cosh\, (\pi \al))~,
\end{equation}
which coincides with the known result.

\subsection{General $p$ and $\kappa$}

Let us start by introducing some more notation:
\begin{equation}
f(\ell) \ := \ \log \left(\frac{\ell+\tfrac{3}{2}+\ii \al }{\ell+\tfrac{1}{2}-\ii \al} \right)~.
\end{equation}
The potentials will be given by the sum of two terms
\begin{equation}
V_{p}^\kappa(\al)  \ = \ U_{p}^\kappa(\al) +U_{p}^{p-\kappa}(-\al) ~.
\end{equation}
By working out some explicit examples, one can convince oneself that the general form of each contribution $U$ is as follows
\begin{equation}
 U_{p}^\kappa(\al) = \sum_{\ell=0}^\infty \left[ s_0 f(p \ell) +s_1 f(p \ell+1) +...+s_{p-1} f(p \ell+(p-1))\right]~,
\end{equation}
where $s_0,...,s_{\ell-1}$ depend on $p, \kappa$ and $\ell$. The important fact is that they are always of the form $s_i = a_i+ b_i \ell$. $V_{p}^\kappa(\al)$ can be computed in two steps. First we use the intermediate result
\begin{eqnarray}
\sum_{\ell=0}^\infty (a + b \ell) \log(p \ell+c +\ii \al) &=& \frac{1}{p} \Big[ (b c- a p+\ii \al b) \log \Gamma \left(\frac{c+\ii \al}{p} \right) \nonumber \\
&& - b p~\psi^{(-2)}\left(\frac{c+\ii \al}{p} \right) \Big]~,
 \end{eqnarray}
where $\psi^{(-2)}\ $ is the polygamma function, and we have dropped a term that can later be fixed (for the final result) by using zeta function regularization.  All our expressions are the sum of such building blocks. In order to assemble the correct building blocks, we just have to compute $s_i=a_i+b_i \ell$ for the fixed value of $p, \kappa$ that we are interested in. Finally, once we have computed the $s_i$, we can compute the correct integration constants by using zeta function regularization, as shown above. It is straightforward to write a Mathematica code that computes the final potential $V_{p}^\kappa(\al)$ for any choice of $p, \kappa$.\footnote{The code is available upon request.} The point is that since $s_i$ is at most linear in $\ell$, we can compute $s_i$ by looking at the terms with $0 \leq \ell \leq 2p-1$.

\subsection{Some explicit examples}

Below we present some explicit results that are used in the numerics. Even though the general answer depends on polygamma functions, for some cases the final expression can be simplified:
\bea
 V_{1}^{0}(\al) & = & -\log \left[2\cosh(\pi\al)\right]~,\nonumber\\
V_{2}^0(\al) &=& -\frac{1}{2} \log [2 \cosh(\al \pi)] + 2\al \cot^{-1} \ex^{\pi \al} +\frac{\ii}{\pi} \left[\mathrm{Li}_2(-\ii \ex^{-\al \pi}) - \mathrm{Li}_2(\ii \ex^{-\al \pi}) \right]~,\nonumber \\
V_{2}^1(\al) &=& -\frac{1}{2} \log (2 \cosh(\al \pi) ) -2\al \cot^{-1} \ex^{\pi \al} -\frac{\ii}{\pi} \left[ \mathrm{Li}_2(-\ii \ex^{-\al \pi}) - \mathrm{Li}_2(\ii \ex^{-\al \pi}) \right]~,\nonumber \\
V_{3}^0(\al) &=& \log \frac{2 \cosh^2(\pi \al/3)}{\cosh \pi \al}~,\nonumber\\
V_{3}^1(\al) &=&V_{3,1}^2(\al) \ = \ -\log [2 \cosh (\pi \al/3)]~.
\eea
A general formula is given in appendix C for the case $\kappa=0$. 
Some comments are in order. We see that the sum of potentials over all $\kappa$ for fixed $p$ satisfies a completeness condition
\begin{equation}
\sum_{\kappa=0}^{p-1} V_{p}^\kappa \ = \ V_1 ~.
\end{equation}
This is of course expected, since fixing $\kappa$ projects over certain terms in the sum giving $V_1$. Another comment is that from the structure of the sums we expect $V_{p}^0(\al)$ to be an even function and $V_{p}^\kappa(\al)=V_{p}^{p-\kappa}(-\al)$.

 \section{A wave of eigenvalues}\label{app:numerics}
 
\begin{figure}[ht!]
\centering
\includegraphics[scale=1.8]{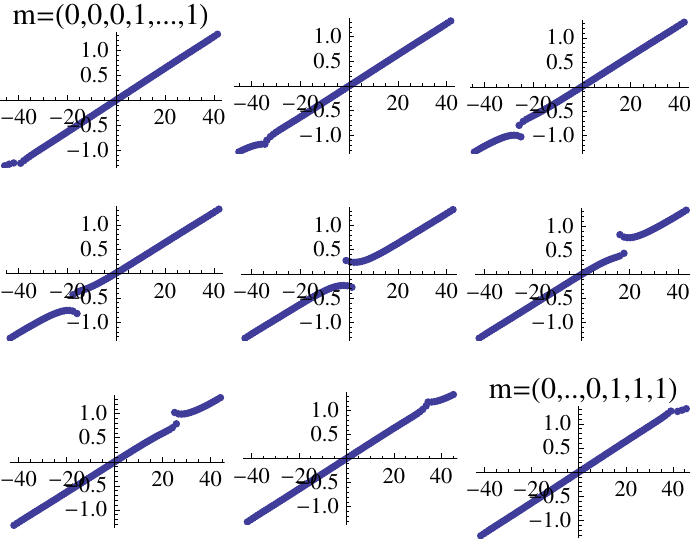}
\caption{Distribution of eigenvalues for $N=100,~p=2$ and $n=3$, $10$, $20$, $30$, $50$, $70$, $80$, $90$, $97$ from left to right and top to bottom.}\label{wave}
\end{figure}

 In the body of the paper we have shown that in the large $N$ limit we can focus on the case ${\bf m} = {\bf \tilde m} = c \cdot 1_{N \times N} $. Furthermore, we have analyzed numerically the distribution of eigenvalues for this case. One can use the numerics to analyze the eigenvalue distribution for other choices of ${\bf m}$ and ${\bf \tilde m} $. These will presumably be important if one wants to compute subleading corrections to our result. An interesting distribution of eigenvalues is obtained if ${\bf m}={\bf \tilde m} \neq c \cdot 1_{N \times N}$ -- see Figure \ref{wave}.
We have shown the eigenvalue distribution for $p=2$ and $N=100$, for the cases ${\bf m} = {\bf \tilde m} = (0, \ldots,0, 1,\ldots, 1)$, with $n$ zeros and $N-n=100-n$ ones. From left to right, top to bottom, we show the eigenvalue distribution for $n=3$, $10$, $20$, $30$, $50$, $70$, $80$, $90$ and $97$. The distribution of eigenvalues is reminiscent of a wave moving from left to right, with the location of the kink at the boundary between the group of zeros and the group of ones. 

\begin{figure}[ht!]
\centering
\includegraphics[scale=0.7]{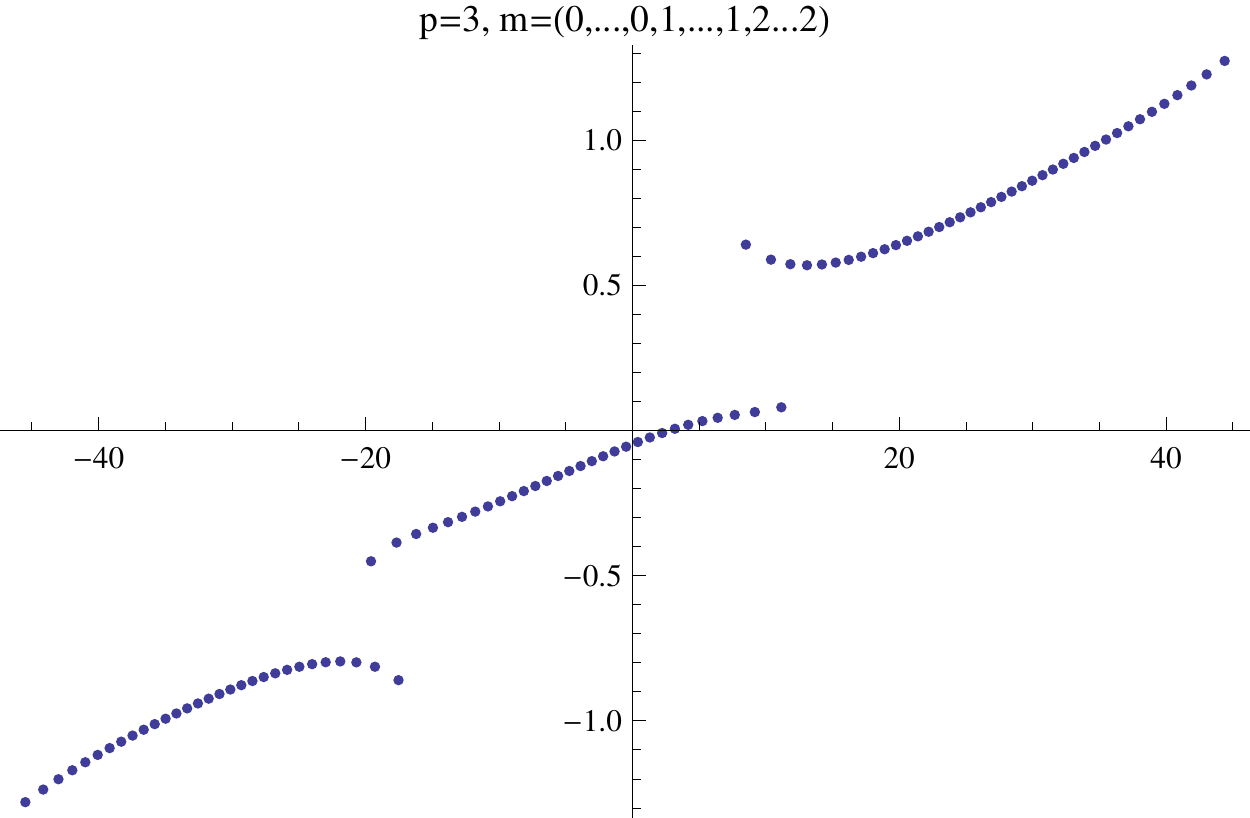}
\caption{Distribution of eigenvalues for $N=100,~p=3$ and ${\bf m} = {\bf \tilde m} = (0,\ldots,0, 1,\ldots, 1, 2,\ldots, 2)$. We see that the eigenvalues distribute in three segments.}\label{anotherwave}
\end{figure}

Equivalently, we see that for $p=2$ and ${\bf m} = {\bf \tilde m}$ the eigenvalues distribute in two segments. The length of the segments is equal to the quantity of zeros and ones, respectively. The numerics seem to suggest that when the number of zeros and ones is ``macroscopic'' ({\it i.e.} of the same order as $N$) there is a finite ``jump'' between the segments. This feature is also present for other cases. For instance, Figure \ref{anotherwave} shows the eigenvalue distribution for $p=3$, $N=100$ and ${\bf m} = {\bf \tilde m} = (0,\ldots,0, 1,\ldots,1, 2,\ldots,2)$, with $30$ zeros, $30$ ones and $40$ twos. It would be interesting to understand whether the finite ``jump'' between segments is really there for large $N$ or an artifact of $N$ being not large enough.

\section{Asymptotic expansions}\label{app:asymptotics}

 In this appendix we present the asymptotic expansions for the potentials found above. We will present a detailed analysis for $\kappa=0$, since, as discussed above, this is enough to compute the free energy in the large $N$ limit.
  
Proceeding as explained in Appendix A we find the following expressions for $p$ odd and even respectively:
\begin{eqnarray}
\partial_\al V_{p=\mathrm{odd}} & = & \frac{\ii}{p} \sum_{n=0}^{p-1} (-1)^n \left[ \psi \left(\frac{1+2n+2 \ii \al}{2p}\right) -  \psi \left(\frac{1+2n-2 \ii \al}{2p}\right) \right]~, \nonumber\\
\partial_\al V_{p=\mathrm{even}} &=&    \frac{1}{p^2} \sum_{n=0}^{p-1} (-1)^n  \Bigg[ (2\al+\ii(2n+1-p)) \psi \left(\frac{1+2n+2 \ii \al}{2p}\right) \nonumber\\
&& + (2\al-\ii(2n+1-p)) \psi \left(\frac{1+2n-2 \ii \al}{2p}\right) \Bigg]~.
\end{eqnarray}
For $p$ odd these expressions can be given in terms of trigonometric functions, but the present form is more uniform. 

Now we would like to compute the asymptotic expansions of such expressions, when the real part of $\al$ is very large. We have the following for $|z| \rightarrow \infty$ and $\arg(z)$ very close to $\pi$:
\begin{equation}
\psi(z) \ = \ \log(z) -\frac{1}{2z} - \sum_{k=1}^\infty \frac{B_{2k}}{2k z^{2k}} + \frac{1}{2} \ii \pi (\ii \cot(\pi z)-1) ~,
\end{equation}
where $B_{2k}$ are the Bernoulli numbers. The coefficient in front of $ \ii \pi (\ii \cot(\pi z)-1) $ is actually one if $|\arg(z)|>\pi$, and is zero otherwise. For the case of real functions, as we are considering, these two average to $1/2$.

It is now easy to compute the asymptotic expansion, substituting the expansion for $\psi$ into the expression for $\partial_\al  V_{p}$. For each case we obtain:

\subsection{$p$ odd}

\begin{equation}
\partial_\al V_{p=\mathrm{odd}} \ = \ -2\frac{\pi}{p} - \frac{2\pi}{p} \sum_{\ell=1}^\infty \frac{\ex^{-2 \frac{\pi}{p} \al \ell }}{\cos(\frac{\pi \ell}{p})}~.
\end{equation}
This gives the following expansion for $ V_{p=\mathrm{odd}}$:
\begin{equation}
 V_{p=\mathrm{odd}} \ = \ -2 \frac{\pi}{p} \al +  \sum_{\ell=1}^\infty \frac{\ex^{-2 \frac{\pi}{p} \al \ell }}{\ell  \cos(\frac{\pi \ell}{p})}~.
\end{equation}
Quite remarkably, zeta function regularization implies a value for $V(\al=0)$ such that the asymptotic expansion doesn't have a constant term.

Given an expansion of the form 
   \begin{equation}
V \ = \ \al \pi \left(c_0 +\sum_{\ell} c_\ell\, \ex^{-\pi \al \ell} \right) +\sum_\ell d_{\ell}\, \ex^{-\pi \al \ell}~, 
  \end{equation}  
 we have seen in the body of the paper that the contribution relevant at large $N$ is
   \begin{equation}
  J\  = \ \sum_{\ell} \left(\frac{c_\ell}{ \ell} y \sin(\ell y)+\frac{c_\ell}{ \ell^2}  \cos(\ell y)+\frac{d_\ell}{\ell} \cos(\ell y) \right)~.
  \end{equation}
   For the present case we have
  \begin{equation}
  J_{p=\mathrm{odd}} \  = \ \frac{p}{2} \sum_{\ell=1}^\infty \frac{\cos(\frac{2\ell}{p}y)}{\ell^2 \cos(\ell \frac{\pi}{p})} \ = \ \frac{2p^2-3}{24p}\pi^2+\frac{y^2}{2p}~,
  \end{equation}
  where we have assumed that $y$ lies in the range $[-\pi/2,\pi/2]$. This is justified by the numerics.
  
  \subsection{$p$ even}
For $p$ even it is convenient to focus on the contribution for a fixed value of $n$ in the general expression for $\partial_\al V_{p,1}$. We obtain
\begin{eqnarray}
\partial_\al V_{p}|_n &= & 2(-1)^n\frac{1+2n-p}{p^2} \pi  -\frac{\pi \al }{p} \sum_{\ell=1}^\infty \frac{4 (-1)^{n} \sin (\frac{\pi(2n+1)}{p} \ell ) }{p} \ex^{-2 \frac{\pi}{p} \al \ell }  \nonumber\\
&&+ \frac{2(-1)^n (1+2n-p) \pi}{p^2} \sum_{\ell=1}^\infty \cos\left(\frac{(2n+1)\pi}{p}\ell\right) \ex^{-2 \frac{\pi}{p} \al \ell }~.
\end{eqnarray}
The contribution from this term to $J$, which we denote $J|_n$, can be computed as above. We obtain
\begin{equation}
J|_n \ = \ (-1)^n \left(-\frac{(1+2n-p)(1+4n^2-4n(p-1)+2p(p-1))}{24p^2} \pi^2 +\frac{p-1-2n}{2p^2} y^2 \right)~.
\end{equation}
Summing over $n$ from zero to $p-1$ and using that $p$ is even we obtain
  \begin{equation}
  J_{p=\mathrm{even}} \ =\  \frac{2p^2-3}{24p}\pi^2+\frac{y^2}{2p}~.
  \end{equation}
This has exactly the same form as for $p$ odd! We thus arrive at the following result, valid for all values of $p$:
  \begin{equation}
  J_{p} \ = \ \frac{2p^2-3}{24p}\pi^2+\frac{y^2}{2p}~.
  \end{equation}

\end{document}